\documentclass[aps, prl, twocolumn, nofootinbib, preprintnumbers]{revtex4-1}%
\pdfoutput=1
\usepackage{graphicx, adjustbox, amsmath, amssymb}

\usepackage[colorlinks=true, citecolor=blue, urlcolor=blue, linkcolor=blue, breaklinks=true, pdfpagelabels=false]{hyperref}
\def\equationautorefname~#1\null{Eq.\,(#1)\null}
\def\sectionautorefname~#1\null{Sec.\,#1\null}
\def\figureautorefname~#1\null{Fig.\,#1\null}

\makeatletter\g@addto@macro\bfseries{\boldmath}\makeatother

\newcommand{\bea}{\begin{eqnarray}}
\newcommand{\eea}{\end{eqnarray}}
\newcommand{\beq}{\begin{equation}}
\newcommand{\eeq}{\end{equation}}

\allowdisplaybreaks[4]

\usepackage{xcolor}
\usepackage[normalem]{ulem}

\begin{document}
\preprint{CERN-TH-2022-012}
\title{Gegenbauer's Twin}

\author{Gauthier Durieux}\email{gauthier.durieux@cern.ch}\affiliation{CERN, Theoretical Physics Department}
\author{Matthew McCullough}\email{matthew.mccullough@cern.ch}\affiliation{CERN, Theoretical Physics Department}
\author{Ennio Salvioni}\email{ennio.salvioni@unipd.it}\affiliation{Universit\`a di Padova, Dipartimento di Fisica e Astronomia and INFN, Sezione di Padova}
\begin{abstract}%
In Twin Higgs models the dominant source of fine-tuning is the cancellation of order $v^2/f^2$ required to obtain a Standard Model-like Higgs, where $v$ and $f$ are the electroweak and new physics scales, respectively.
Recently proposed Gegenbauer Goldstone models naturally realise $v^2/f^2 \ll 1$ and hence remove this source of fine-tuning.
By combining the two into `Gegenbauer's Twin', we obtain a symmetry-based model for Higgs-sector naturalness consistent with current collider measurements without fine-tuning of parameters.
Single-Higgs coupling deviations of a few percent and trilinear self-coupling deviations of order one are irreducible in the natural parameter space.
Thus, notably, the fingerprints of Gegenbauer's Twin could emerge first through di-Higgs measurements at the High-Luminosity LHC.
\end{abstract}

\maketitle

\section{Introduction}

It is beyond doubt that the Standard Model (SM) should be replaced by a more fundamental description at some high energy scale $\Lambda$.
If the Higgs mass becomes a physical quantity calculable in terms of the parameters of that more fundamental theory, as one would expect for instance if the Higgs arises as a pseudo Nambu-Goldstone boson (pNGB), then the question of the $m_h \ll \Lambda$ scale separation arises. In many theories, this separation can only be achieved by fine-tuning parameters.

By introducing a hidden copy of the SM, related to it by a $\mathcal{Z}_2$ exchange symmetry, Twin Higgs models~\cite{Chacko:2005pe,Barbieri:2005ri,Chacko:2005vw,Chacko:2005un} go some way towards alleviating this tension.
The new particles that protect the pNGB Higgs mass from large quantum corrections do not interact through the gauge and Yukawa forces of the SM, and thus can be rather light.
However, these models naturally predict either a vanishing electroweak scale $v$, or no separation between the electroweak and Twin breaking scales, $v \sim f$.
As such, in all existing realisations an additional source of exchange symmetry breaking is introduced which allows one to obtain $v \ll f$ at the price of a fine-tuning of magnitude $2 v^2/f^2$.
Since Higgs coupling modifications scale as $v^2/(2 f^2)$, the increasingly precise LHC Higgs measurements inevitably necessitate a residual fine-tuning at the $20\%$ level or worse.%
\footnote{Models with tadpole-induced electroweak symmetry breaking~\cite{Harnik:2016koz} are an interesting exception, but their compatibility with LHC direct searches and electroweak precision tests is currently unclear~\cite{Contino:2017moj}.}

Recently, a new class of explicit symmetry breaking operators for pNGBs have been introduced, wherein a spurion taking values in an irreducible representation of the global symmetry gives rise at low energies to a radiatively stable Gegenbauer polynomial potential~\cite{Durieux:2021riy}.
Owing to the structural features of these functions, a parametric separation $v \ll f$ is naturally obtained.
However, application to classic models of composite pNGB Higgs in \cite{Durieux:2021riy} showed that the LHC lower bounds on coloured top partner masses entail some degree of fine-tuning.
It is thus tempting to speculate, as in~\cite{Durieux:2021riy}, that combining a Twin Higgs model with a Gegenbauer potential may lead to {\it fully natural} electroweak symmetry breaking (EWSB).
In this paper, we demonstrate that this is indeed the case.

We first generalise the construction of~\cite{Durieux:2021riy} to the explicit symmetry breaking pattern relevant to the Twin Higgs, namely $\text{SO}(2N) \to \text{SO}(N)\times \text{SO}(N)$, deriving the structure of resultant Gegenbauer potentials.
Then we focus on a concrete model, inspired by the composite Twin Higgs of~\cite{Barbieri:2015lqa}, to quantitatively evaluate the fine-tuning and demonstrate that a fully natural theory is obtained for $f \sim \mathrm{TeV}$.
Finally, we highlight its leading phenomenological prospects, which include $\mathcal{O}(1)$ modifications of the Higgs trilinear self-coupling, a tantalising prediction that will be tested by the High-Luminosity LHC.
In the appendix, we include a self-contained discussion on the tight connection between radiative stability and naturalness.

\section{Gegenbauer's Twin}
\label{sec:GT}
The original Twin Higgs model~\cite{Chacko:2005pe} extended the SM by an exact mirror (``Twin'') copy $\text{SM}_\text{T}$.
The only interaction between the two sectors was through the Higgs potential, assumed to take the $\text{SO}(8)$-invariant form
\beq
V= \lambda \left( |H|^2+|H_T|^2-f^2/2 \right)^2 ~~,
\label{eq:potlead}
\eeq
and causing the spontaneous $\text{SO}(8)\to \text{SO}(7)$ breaking.
Following the spirit of \cite{Durieux:2021riy}, we focus here on the infrared (IR) structure, leaving open the question of the ultraviolet (UV) completion, which may be composite~\cite{Barbieri:2015lqa,Low:2015nqa,Contino:2017moj}, contain extra dimensions~\cite{Geller:2014kta,Craig:2014aea,Craig:2014roa}, or involve supersymmetry~\cite{Falkowski:2006qq,Chang:2006ra,Craig:2013fga,Katz:2016wtw,Badziak:2017syq,Badziak:2017kjk,Badziak:2017wxn}.
For the purpose of understanding the IR structure of the theory, we may package the $8$ real scalar degrees of freedom into an $\mathbf{8}$ of $\text{SO}(8)$ denoted $\boldsymbol{\omega} = (f+\rho) \boldsymbol{\phi}\,$. Here $\rho$ is the radial mode of the spontaneous symmetry breaking and $\boldsymbol{\phi}$ parameterises the vacuum manifold $\boldsymbol{\phi}\cdot \boldsymbol{\phi} =1$, 
\beq
\boldsymbol{\phi} =\frac{1}{\Pi} \sin \frac{\Pi}{f} \begin{pmatrix}
           \Pi_{1} \\
           \vdots \\
           \Pi_{2N-1} \\
          \Pi \cot \frac{\Pi}{f}
         \end{pmatrix}\;,
         \qquad\text{with } \Pi = \sqrt{\boldsymbol{\Pi} \cdot \boldsymbol{\Pi}}~~~,
\eeq
where we have generalised $\text{SO}(8)$ to $\text{SO}(2N)$.
The first $N$ components of $\boldsymbol{\omega}$ would comprise the Higgs multiplet and the latter $N$ the Twin Higgs.

We now construct the traceless symmetric irreps that explicitly break $\text{SO}(2 N)\to \text{SO}(N) \times \text{SO}(N)$. For the sake of generality we retain the radial mode $\rho$ in the discussion, although to analyse the vacuum structure we later focus on the effective theory below its mass.
Defining the spurion $\widetilde{D} = \mathrm{diag}\,( - {\bf 1}_N, + {\bf 1}_N)$, which has formal transformation property $\widetilde{D} \to R \widetilde{D} R^T$ under SO$(2N)$, the desired irreps may be found from the Taylor expansion
\beq
\begin{aligned}
F (t \boldsymbol{\omega})
&\equiv \left(1
	- 2t^2 \boldsymbol{\omega}^T \widetilde{D}\, \boldsymbol{\omega}
	+ t^4 (\boldsymbol{\omega} \cdot \boldsymbol{\omega})^2  \right)^{(1-N)/2}
\\&= \sum_{n=0}^\infty t^{2n} K_{2n}^{i_1i_2...i_{2n}} \omega_{i_1} \dots \omega_{i_{2n}} ~~.
\end{aligned}
\label{eq:generating-fun}
\eeq
The tensors $K_{2n}$, given by
\bea
K_{2n}^{\,i_1 \ldots i_{2n}}   \equiv  \frac{1}{(2n)!} \frac{\partial^{2n} F(\boldsymbol{\hat{\phi}})}{\partial \hat{\phi}_{i_1} \ldots \partial \hat{\phi}_{i_{2n}}}  \bigg|_{\boldsymbol{\hat{\phi}} \,=\, 0}~~,
\label{eq:irrepdef}
\eea
are manifestly symmetric and also traceless, as can be verified by making use of the properties $\mathrm{Tr}\,\widetilde{D} = 0$ and $\widetilde{D}^2 = {\bf{1}}_{2N}$. Moreover, expressing the first line of \autoref{eq:generating-fun} as $\big(1
- 2\hat{t}^{\,2} \boldsymbol{\phi}^T \widetilde{D}\, \boldsymbol{\phi}
+ \hat{t}^{\,4} \big)^{(1-N)/2}$ with $\hat{t}\equiv t(f+\rho)$, one recognises the generating function of Gegenbauer polynomials $G_n^{\nu}$ with $\nu = (N-1)/2$.
Therefore, by identifying each order in $\hat{t}$ one finds 
\bea
K_{2n}^{\,i_1 \ldots i_{2n}}  \phi_{i_1} \ldots \phi_{i_{2n}} =&\; G_n^{(N-1)/2} \left( \boldsymbol{\phi}^T \widetilde{D}\, \boldsymbol{\phi} \right)~~,
\eea
implying that the explicit breaking of $\text{SO}(2 N)\to \text{SO}(N) \times \text{SO}(N)$ by a traceless symmetric irrep leads to a potential taking the form of a Gegenbauer polynomial.

Hence, introducing a small dimensionless parameter $\epsilon$ and the radial mode mass  $m_\rho=\sqrt{2\lambda} f$, we may identify any pNGB potential of the form
\beq
V_G^{(n)} = \epsilon f^2 m_\rho^2  (1+\rho/f)^{2 n} G_n^{(N-1)/2} \left(  \boldsymbol{\phi}^T \widetilde{D}\, \boldsymbol{\phi} \right) ~~,
\label{eq:geg-pot}
\eeq
as being radiatively stable against UV corrections, at $\mathcal{O}(\epsilon)$ and all loop orders.
This is because it arises from an explicit symmetry-breaking UV spurion sitting in an irrep of $\text{SO}(2N)$, which is traceless and symmetric.
As a result, UV corrections may multiplicatively renormalise a $V_G^{(n)}$ potential but will not alter its functional form at $\mathcal{O}(\epsilon)$.

The radiative stability can also be seen from a one-loop Coleman-Weinberg (CW) calculation below the radial mode mass.
Assuming the pNGB potential to be a function $V(x)$ of $x \equiv \boldsymbol{\phi}^T \widetilde{D}\, \boldsymbol{\phi}$, the leading, quadratically divergent piece of the CW is found to be
\beq
V_{\rm CW}^{\Lambda^2} = \frac{\Lambda^2}{8\pi^2 f^2} \left[ (1 - x^2) \frac{\partial^2 }{\partial x^2} - N x \frac{\partial }{\partial x} \right]\,V(x).
\eeq
Since $G_n^{(N-1)/2}(x)$ is an eigenfunction of this differential operator, the multiplicative renormalisation of $V_G^{(n)}$ at one loop and linear order is confirmed.
Both perspectives were elucidated further in~\cite{Durieux:2021riy}, for a slightly different explicit symmetry breaking pattern.

\subsection{Vacuum Structure}
We now turn to a discussion of the vacuum structure of Gegenbauer's Twin, focussing on the effective theory for the pNGBs below the radial mode mass.
Therefore, the following analysis applies directly to composite realisations~\cite{Barbieri:2015lqa,Low:2015nqa,Contino:2017moj,Geller:2014kta}, where $\lambda$ is effectively large.
We expect the main qualitative features will also apply to weakly coupled supersymmetric completions~\cite{Falkowski:2006qq,Chang:2006ra,Craig:2013fga,Katz:2016wtw}, but quantitative differences may arise due to the lightness of $\rho$ and the presence of a second Higgs doublet, as required by holomorphy of the superpotential.

The gauging of the SM and Twin electroweak (EW) groups leads to $6$ pNGBs being eaten by massive gauge bosons, leaving only the Higgs field $h$ as physical scalar degree of freedom, as manifest in the unitary gauge where $\Pi_{i} = \delta_{i4} h$ and $\boldsymbol{\phi}^T \widetilde{D}\, \boldsymbol{\phi} = \cos (2h/f)$.
The explicit $\text{SO}(8)$ breakings introduced by the EW gauge and fermion Yukawa interactions generate a perturbatively estimable potential for $h$.
If the couplings are symmetric under the $\mathcal{Z}_2$ exchange acting as $h/f \leftrightarrow \pi/2 - h/f$, this is dominated by (see e.g.~\cite{Craig:2015pha,Barbieri:2015lqa,Low:2015nqa})
\beq\label{eq:V_SM}
V_t \approx \beta f^4  \bigg[ \sin^4 \frac{h}{f}\, \log \frac{a}{\sin^2 \frac{h}{f} } + \cos^4 \frac{h}{f}\, \log \frac{a}{\cos^2 \frac{h}{f} } \bigg] ~~,
\eeq
where $\beta = {3 y_t^4}/{(64 \pi^2)}$ with $y_t = y_t^{\rm \overline{MS}} (m_t) \approx 0.94$ and $\log a$ is an $\mathcal{O}(1)$ quantity.
Note that the $a$-dependent threshold and UV contributions to the potential which are not logarithmically dependent on the Higgs field take precisely the functional form of a Gegenbauer polynomial $G_n^{3/2}(\cos 2h/f)$ with $n=2$,%
\footnote{Indeed, $\sin^4h/f + \cos^4 h/f$ and $\cos^2 2h/f$ are equivalent modulo an $\text{SO}(8)$-invariant constant.}
as is to be expected on general grounds for a radiatively stable contribution.
The potential in \autoref{eq:V_SM} is characterised by three parameter regions with different symmetry-breaking patterns~\cite{Barbieri:2015lqa}, depending on the value of $\log a$ (namely $\log a$ below, between, or above $1/2$ and $3/2-\log2$).
In isolation, none of these possibilities is however realistic.

\begin{figure}
\includegraphics[width=1\columnwidth]{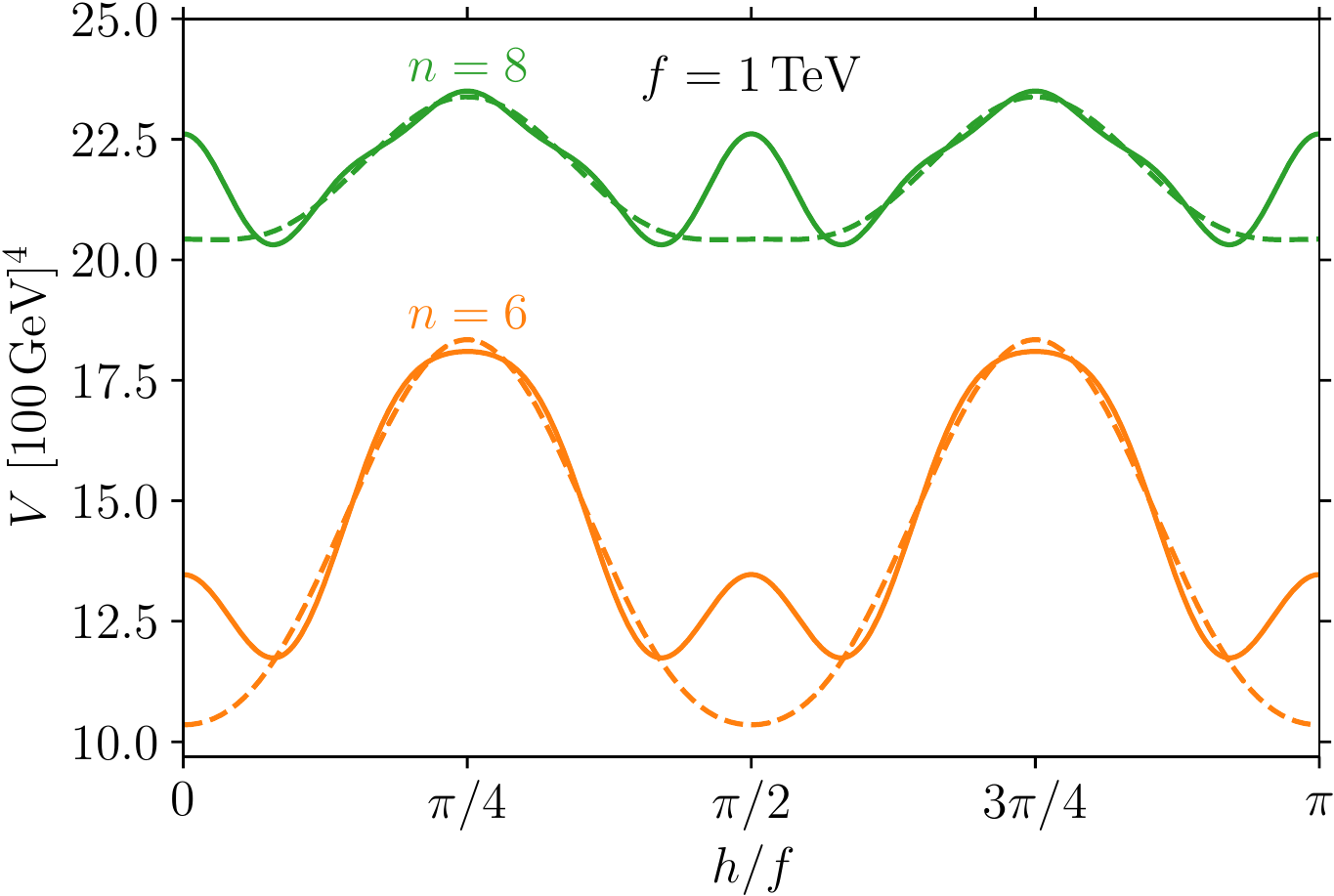}
\caption{%
Full Gegenbauer's Twin potential $V=V_t+V_G^{(n)}$ for $n=6, 8$ and $f=1\,$TeV.
Dashed lines show the top-sector component $V_t$.
Parameters are fixed so as to reproduce the physical Higgs mass and vev in the first minimum.
}
\label{fig:potential}
\end{figure}

\begin{figure*}[t]
\adjustbox{max width=1\textwidth}{%
\includegraphics{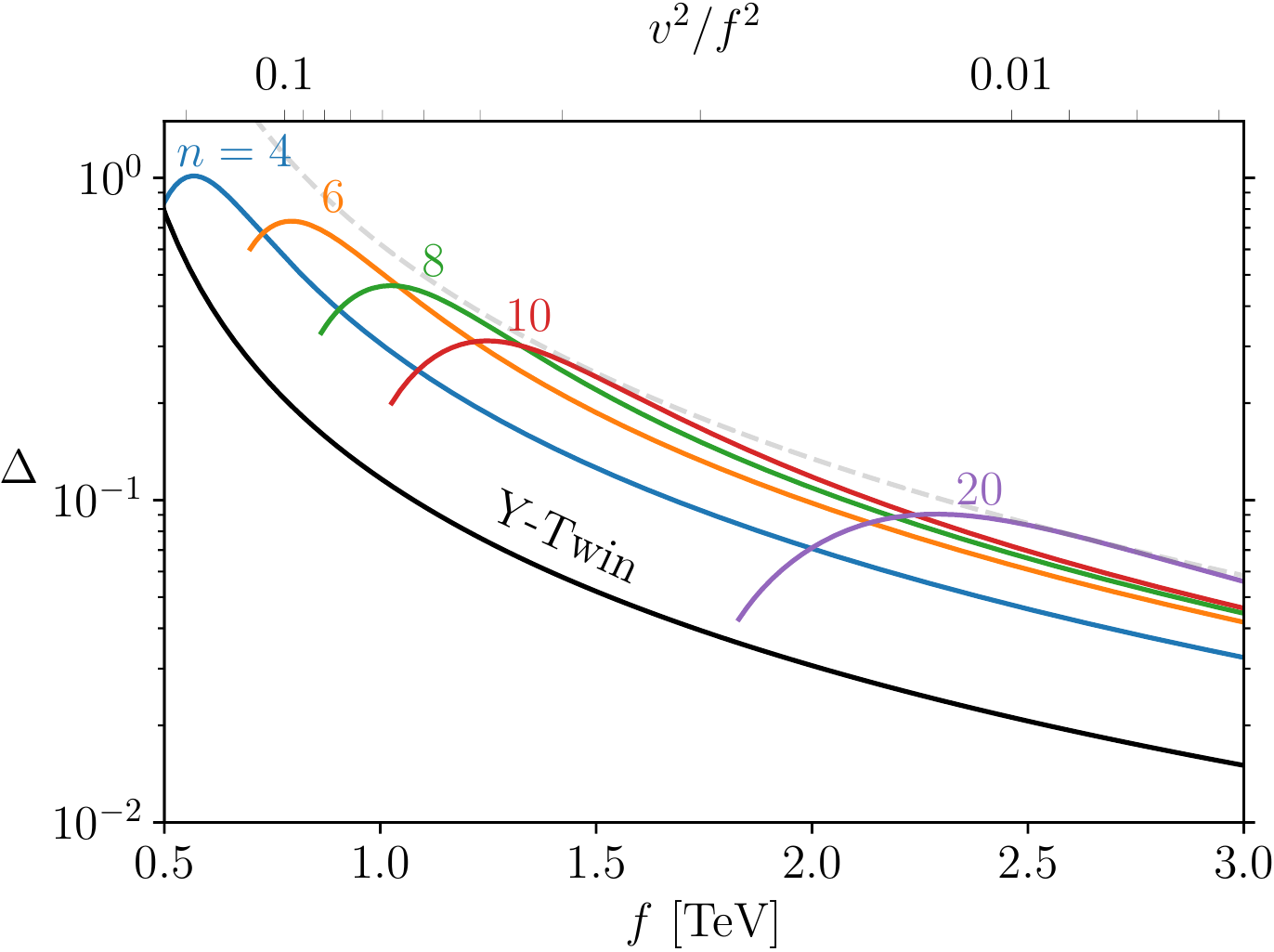}\hspace{3em}%
\includegraphics{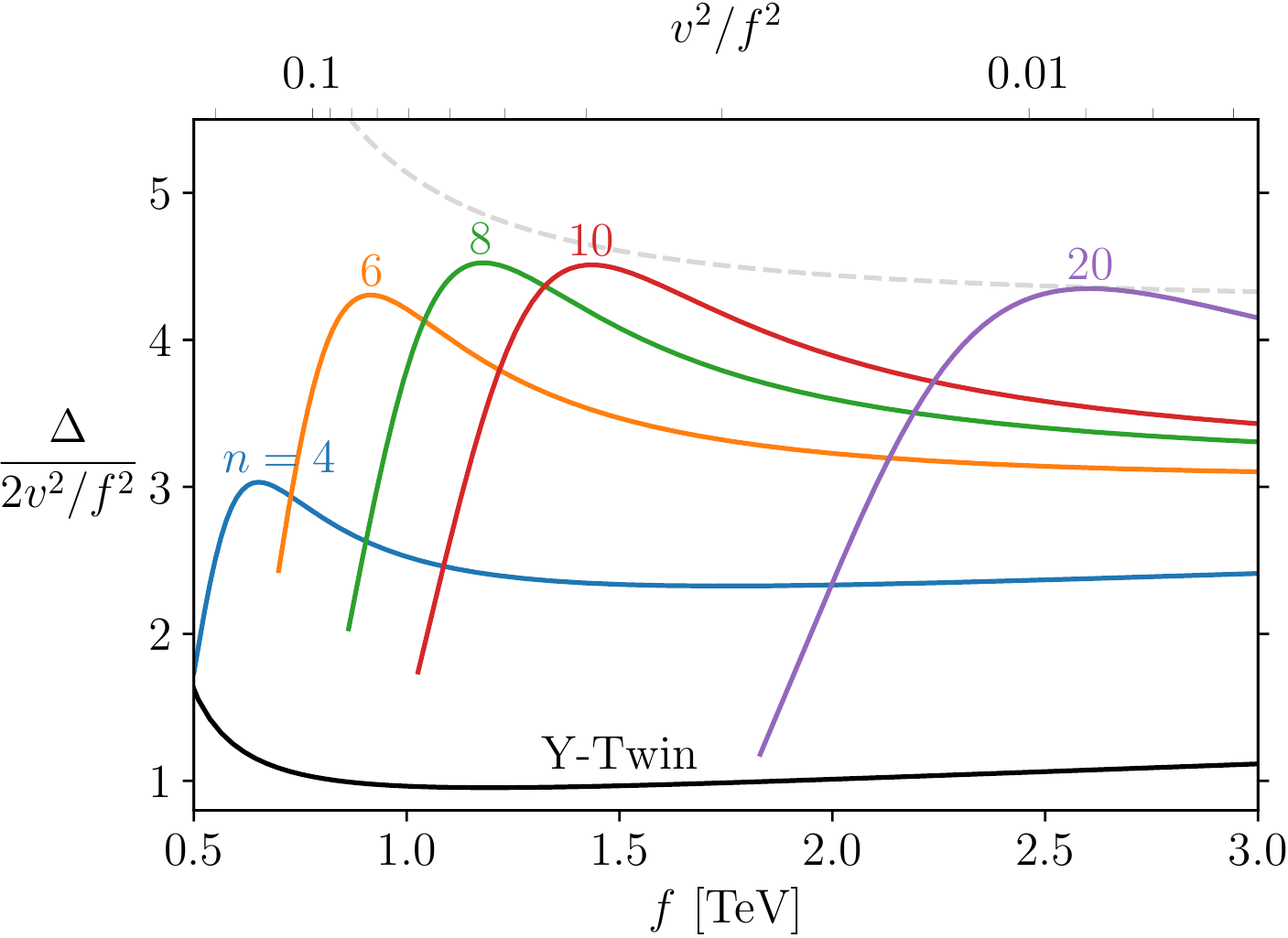}%
}%
\caption{\label{fig:tuning}%
{\it (Left)} Total fine-tuning of Gegenbauer's Twin for different spurion index $n$, as a function of the symmetry breaking scale $f$.
For comparison, we also show the composite Twin Higgs model of \cite{Barbieri:2015lqa} (denoted Y-Twin), where the $\mathcal{Z}_2$ breaking arises from not gauging the Twin hypercharge.
The dashed line shows the $n$-independent $({\partial \log  v^2}/{\partial \log  a})^{-1}$ contribution that dominates the tuning of Gegenbauer's Twin.
{\it (Right)} Ratio of the tuning of Gegenbauer's Twin and Y-Twin models to the naive $2 v^2/f^2$ estimate expected for the latter.
}
\end{figure*}

To obtain viable EWSB, we introduce a new UV source of explicit symmetry breaking.
We assume the presence of a spurion in the $n$-index irreducible representation of SO$(8)$, whose physical expectation value causes an explicit $\text{SO}(8)\to \text{SO}(4)\times \text{SO}(4)$ breaking.
In a strongly coupled UV completion, this spurion may be thought of as having an origin within the composite sector.
In the unitary gauge, it generates a contribution to the Higgs potential of the form
\beq\label{eq:V_G}
V_G^{(n)} =  \epsilon\, m_\rho^2 f^2 G_n^{3/2} \left( \cos 2 h/ f \right) ~~.
\eeq
For odd $n$, the Gegenbauer potential $V_G^{(n)}$ is minimised at $\langle h \rangle/f = \pi/2$, resulting in $v = f$, which is experimentally ruled out.
For even $n$, the full potential $V = V_t + V_G^{(n)}$ (see \autoref{fig:potential}) is exactly $\mathcal{Z}_2$ symmetric.
It results in spontaneous $\mathcal{Z}_2$ breaking%
\footnote{See e.g.~\cite{Beauchesne:2015lva,Harnik:2016koz,Yu:2016swa,Batell:2019ptb,Csaki:2019qgb} for other Twin Higgs models where the $\mathcal{Z}_2$ exchange symmetry is not broken explicitly.}
and realistic EWSB, with a preference for the region $ \log a < 1/2$ where $V_t$ alone has minima at $h/f = 0, \pi/2$.
Given $\{f, n\}$ inputs, we determine the $\{{\epsilon}, \log a\}$ parameters that yield the observed values for $\{v, m_h\}$, where $v = f \sin ( \langle h \rangle / f) \approx 246\;\mathrm{GeV}$ with $\langle h \rangle$ the location of the first, deepest, minimum of the potential, and $m_h = 125\;\mathrm{GeV}$.
For instance, for $f = 1\;\mathrm{TeV}$ and $n=6$, we find  $\epsilon\, m_\rho^2/f^2 \approx 1.1\times 10^{-5}$ and $\log a \approx 0.27$.
As long as $\log a \lesssim 0.7\,$, with mild $n$ dependence, the first minimum remains the global one.
Henceforth, we only discuss parameter space where this is verified.

As a useful reference, we compare our results to the model of~\cite{Barbieri:2015lqa}, where an explicit $\mathcal{Z}_2$  breaking was introduced by not gauging the Twin hypercharge.
This generates the additional potential $V_Y = \alpha f^4 \sin^2(h/f)$ on top of $V_t$, enabling viable EWSB provided $\log a$ takes larger values compared to our setup, $\log a \approx 6 - \log ( f/v )$.
We term that scenario `hypercharge-breaking Twin' or `Y-Twin' for brevity.
We note that an explicit calculation of $\log a$ in a concrete composite model, albeit still logarithmically sensitive to physics at the cutoff, was provided in Appendix C of \cite{Contino:2017moj}.
The parametric freedom found there, which allows for a negative UV contribution, illustrates the possibility of obtaining values of $\log a$ much smaller than those considered in~\cite{Barbieri:2015lqa}, so that an embedding of our model in the composite Twin Higgs context appears plausible.
Accordingly, in this work we simply take $\log a$ to be an $\mathcal{O}(1)$ parameter, assuming that a natural UV completion exists.

\subsection{Fine-Tuning}
For fixed symmetry breaking scale $f$ and representation index $n$, the fine-tuning is calculated from its log-derivative definition~\cite{Barbieri:1987fn}.
We construct the matrix
\beq
\delta =\begin{pmatrix}
          \frac{\partial \log v^2}{\partial \log  {\epsilon}} & \frac{\partial \log  v^2}{\partial \log  a} \\
          \frac{\partial \log  m_h^2}{\partial \log  {\epsilon}} & \frac{\partial \log  m_h^2}{\partial \log a}
         \end{pmatrix} ~~,
\eeq
which determines the rate of change of the physical observables $v^2$ and $m_h^2$ with respect to variations in the underlying model parameters.
Thus, large entries in this matrix signal large sensitivities.
As a measure of the total tuning, we take
\beq
\Delta = \left( \sum \mathrm{eigenvalues}\, (\delta^T \delta) \right)^{-1/2}~~.
\eeq
Compared to other common definitions based on the inverse of the modulus of the individual entries of $\delta$, ours turns out to be conservative.
For the $V_t + V_Y$ potential considered in~\cite{Barbieri:2015lqa} (with $\alpha$ replacing ${\epsilon}$ in the matrix $\delta$), it yields the expected $\Delta \approx 2v^2 /f^2$ result (see for example~\cite{Craig:2015pha}).

The fine-tuning of Gegenbauer's Twin is dominated by the sensitivity of $v$ and $m_h$ to $\log a$.
For the vacuum expectation value (vev), we find
\beq
\left( \frac{\partial \log  v^2}{\partial \log  a} \right)^{-1} = \frac{8\pi^2 m_h^2}{3 y_t^4 f^2 \big(1 - \frac{3v^2}{f^2} + \frac{2v^4}{f^4} \big) }~~,
\label{eq:dlog2dloga}
\eeq
a result that does not depend on $n$ (and applies to the Y-Twin, as well).
If this is the dominant source, as it is the case for values of $n$ minimising the tuning, then
\beq
\frac{\Delta}{2v^2/f^2} \approx \frac{4\pi^2 m_h^2}{3y_t^4 v^2}\approx 4
\eeq
up to about $25\%$ corrections.
For $f = 1\;\mathrm{TeV}$, $n =6$ or $8$ are optimal and $\Delta \approx 0.5$, namely {\it no tuning}.
These features are illustrated in \autoref{fig:tuning}.

\subsection{Phenomenology}
\begin{figure*}[t]
\adjustbox{max width=1\textwidth}{%
\includegraphics{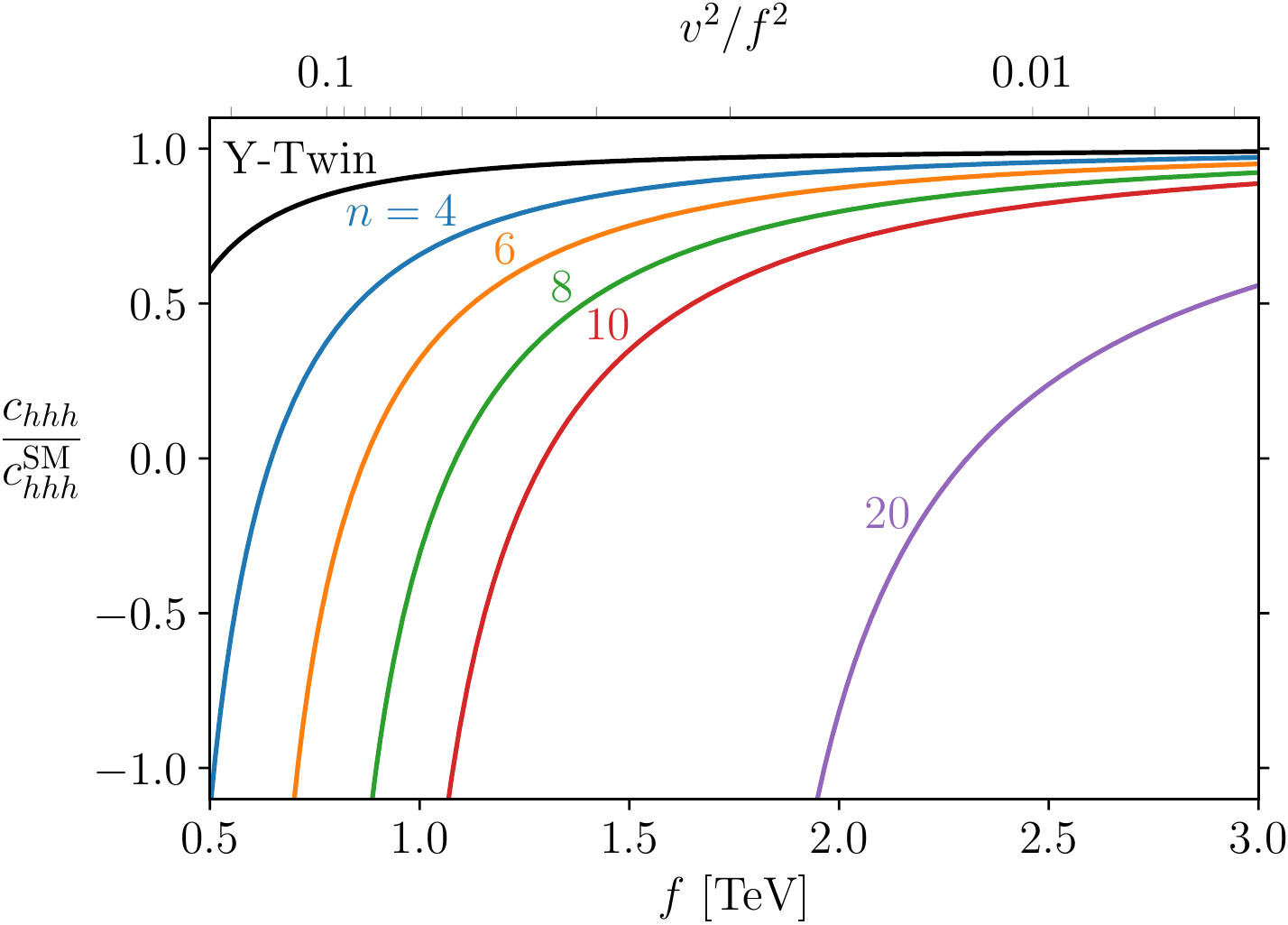}\hspace{3em}%
\includegraphics{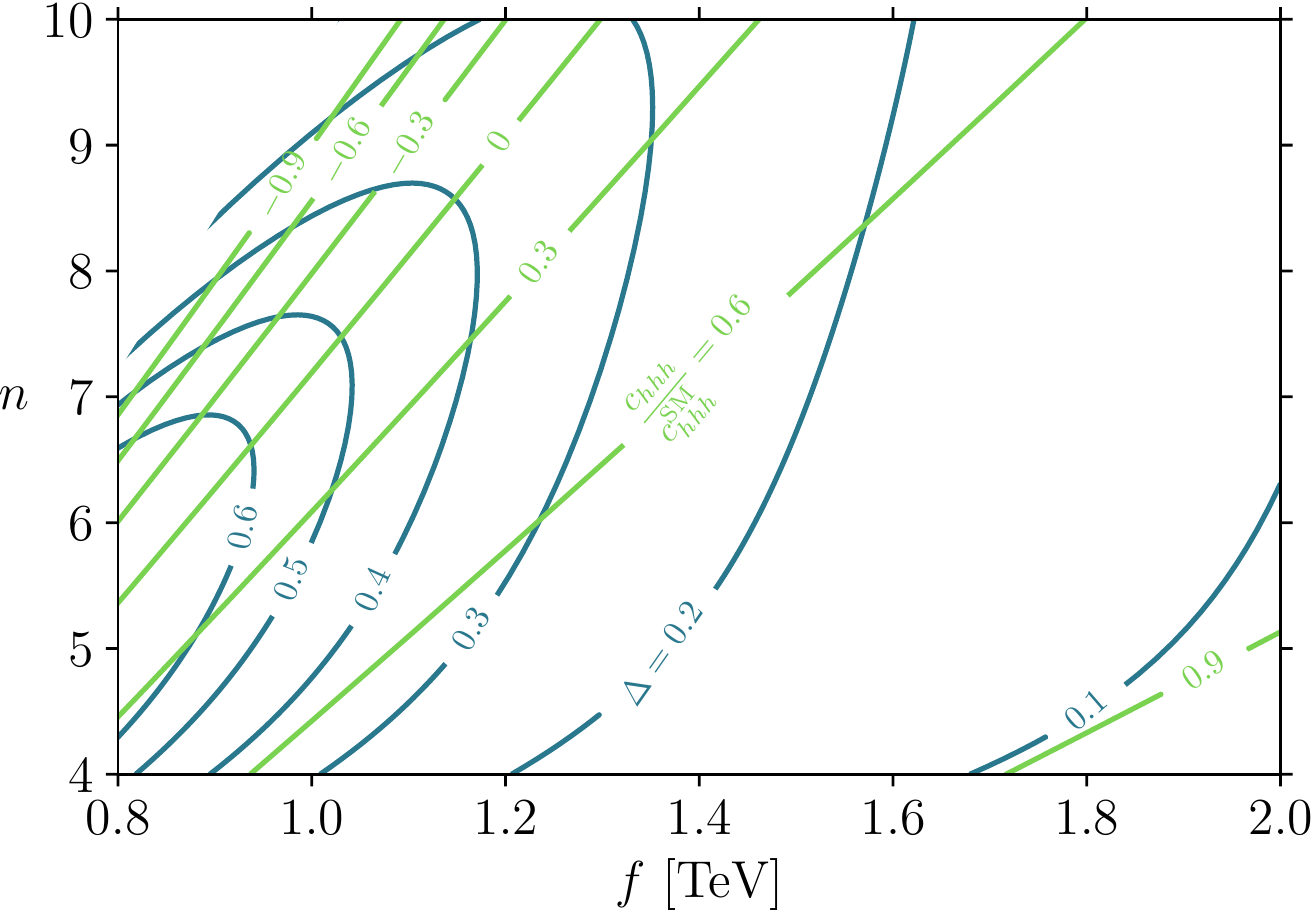}%
}%
\caption{\label{fig:trilinear}%
{\it (Left)} Ratio of the Higgs trilinear self-coupling to its SM expectation.
{\it (Right)} Contours indicating the total fine-tuning $\Delta$ and the $c_{hhh}/c_{hhh}^\text{SM}$ coupling deviation in the plane of input parameters $\{ f, n \}$.
We do not consider the region in the upper-left corner where the first minimum is not the global one.
}
\end{figure*}

In addition to the universal rescaling of single-Higgs couplings by $\sqrt{1 - v^2/f^2}\,$, Gegenbauer's Twin exhibits significant corrections to the Higgs trilinear self-coupling.
These are shown in the left panel of \autoref{fig:trilinear}, normalised to the SM prediction, accounting for the leading one-loop correction arising from top triangle diagrams~\cite{Hollik:2001px} included in $V_t$.
The deviations from the SM are much larger than for standard Twin Higgs models: for $f = 1\;\mathrm{TeV}$, we find $c_{hhh}/c_{hhh}^{\rm SM} \approx +\, 0.32~(-\, 0.31)$ for $n = 6~(8)$, to be compared with $+\,0.91$ for the Y-Twin.
Such large deviations may be visible at the High-Luminosity LHC~\cite{Cepeda:2019klc}. As can be seen in the right panel of \autoref{fig:trilinear}, $\mathcal{O}(1)$ deviations in $c_{hhh}$ are present in all the natural parameter space.
Furthermore, it is conceivable that $c_{hhh}$ would be the first Higgs coupling to show a departure from the SM at colliders. 

The quadrilinear self-coupling also displays large deviations: for $f = 1\;\mathrm{TeV}$, one finds $c_{hhhh}/c_{hhhh}^{\rm SM} \approx -\, 3.1$ $(-\, 4.9)$ for $n = 6~(8)$. The correlation between $c_{hhh}$ and $c_{hhhh}$ may allow future colliders to test the Gegenbauer nature of the Higgs potential (see~\cite{Maltoni:2018ttu,Bizon:2018syu,Chiesa:2020awd} for recent studies).

In our analysis of the EWSB, we have assumed an exact exchange symmetry in the gauge and matter sectors, so that the potential arising from gauge and fermion loops is dominated by \autoref{eq:V_SM}.
Exactly $\mathcal{Z}_2$ symmetric gauge couplings,
including for the SM and Twin hypercharge groups, imply the presence of a massless dark photon in the spectrum.
If the exchange symmetry is not broken explicitly, a {\it mirror Twin Higgs} scenario is realised, which is known to conflict with observations because of a large contribution to the effective number of neutrino species, $\Delta N_{\rm eff} \approx 6$ from the Twin photon and neutrinos~\cite{Chacko:2016hvu}.
This can be resolved by an asymmetric reheating process~\cite{Chacko:2016hvu,Craig:2016lyx}, with interesting predictions for cosmological observables~\cite{Chacko:2018vss}.\footnote{Other possibilities include $\mathcal{Z}_2$ breaking in the neutrino sector~\cite{Csaki:2017spo} or in the Twin Yukawas~\cite{Barbieri:2016zxn,Harigaya:2019shz}.
The latter scenario also generates a one-loop $\mu^2 |H|^2$ term, which is not included in our analysis of the potential.}

In addition to the already discussed $\mathcal{Z}_2$ exchanging the SM and Twin sectors, our potentials exactly preserve a second $\mathcal{Z}_2$ acting as $h/f \leftrightarrow \pi - h/f$.
As a result, the minimum near $h = 0$ is exactly degenerate with three other minima in the interval $0 \leq h/f \leq \pi$.
In the early Universe thermal corrections lift the two minima near $h/f = \pi/2$, but the degeneracy with the minimum near $h/f = \pi$ remains, leading to the appearance of domain walls which may dominate the cosmological energy density.
This issue can however be avoided by introducing a tiny breaking that raises the $h/f \sim \pi$ minimum~\cite{DiLuzio:2019wsw}.

\section{Conclusions}
Through the `Gegenbauer's Twin' model proposed here, we have demonstrated that the commonly accepted $2v^2/f^2$ fine-tuning of Twin Higgs models is the result of a minimality assumption imposed on sources of explicit symmetry breaking, rather than an irreducible effective field theory constraint. Some explicit $\text{SO}(8)$ breaking source is a requirement for any Twin Higgs model. However, if it comes in the form of a higher dimensional irrep, rather than the usual one- or two-index irreps, the `$v/f$' tuning may be eliminated given current collider constraints.

The implications of this work go beyond Twin Higgs models.
The apparent failure of symmetry-based approaches to naturally accommodate the observed separation between the electroweak and UV completion scales has led to speculations about a `naturalness crisis' in particle physics.
As a strictly symmetry-based approach, Gegenbauer's Twin contradicts this hypothesis, suggesting that the crisis may not be with symmetry or effective field theory, but instead with more \ae{}sthetic `minimality' criteria regarding the nature of symmetry-breaking parameters, specifically with regard to pNGB Higgs models.
On the other hand, the IR theory described here offers no explanation as to why the leading explicit symmetry breaking parameters would arise in higher dimensional irreps of global symmetries.
Unless some motivation for this can be found, the overall status of symmetry-based approaches to Higgs naturalness remains unclear.

\begin{acknowledgments}
\vspace{3mm}\noindent{\it Acknowledgments}
We are grateful to Gian Giudice and Alex Pomarol for conversations.
ES acknowledges partial support from the EU's Horizon 2020 programme under the MSCA grant agreement 860881-HIDDeN.
\end{acknowledgments}

\appendix
\section*{Appendix: Why Radiative Stability?}

Radiative stability and fine-tuning are two sides of the same coin. As applied to the Higgs field, a natural theory is one in which the expectation value and mass are \emph{calculable} and \emph{radiatively stable across scales}.
To see how this feeds into a requirement on the nature of the potential, we may use a perturbative Twin toy example.
Since irrep spurions form a complete set, any UV contribution to the pNGB potential preserving $\text{SO}(4)\times \text{SO}(4)$ may be written as the tower of higher dimension operators
\beq
\begin{aligned}
V &= \epsilon f^2 m_\rho^2 \sum_{n = 0}^{\infty} a_n f^{-2n} K_{2n}^{i_1...i_{n}} \omega_{i_1} \dots\, \omega_{i_{2n}}
\\&= \epsilon f^2 m_\rho^2 \sum_{n = 0}^{\infty} a_n  (1+\rho/f)^{2 n} G_n^{3/2} \left(  \cos 2 h/ f \right) ~~,
\end{aligned}
\label{eq:higherdims}
\eeq
where the traceless symmetric tensors $K_{2n}$ were defined in \autoref{eq:irrepdef},
and $\rho$ is the radial mode from \autoref{eq:potlead}.

Note, however, that the symmetric part of the potential
\beq
V= \frac{ \lambda}{4} \left( \boldsymbol{\omega}\cdot\boldsymbol{\omega} - f^2 \right)^2 ~~,
\eeq
also contributes to the renormalisation of the higher dimension operators.
As a result, above the scale of the radial mode the coefficients $a_n$ all run multiplicatively and differently at $\mathcal{O}(\epsilon)$, with $\beta_{a_n} \propto a_n \lambda n^2/4 \pi^2$ at the leading order in $n$ and $\lambda$.
Since the $\beta$-function contribution from the quartic interaction is positive, the higher-$n$ Wilson coefficients decrease more rapidly in running from the UV towards $m_\rho$.
As a consequence, in the UV theory, neither the relative magnitudes of the $a_n$ nor any particularly special linear combination of them are renormalisation-group (RG) invariant.

We may illustrate the essence of this point further with an explicit example.
Suppose that, by hand, we were to postulate a specific form of IR scalar potential mimicking the usual $\text{U}(1)$ case for which a small vev and mass appear plausible within the IR theory
\beq
V = \epsilon f^2 m_\rho^2 \cos \left(  2 k h/ f \right) ~~,
\label{eq:cosans}
\eeq
where $k\in 2 \mathbb{Z}_+$.
There is a minimum at $h/f=\pi/(2k)$, hence for large $k$ and small $\epsilon$ the vev and mass may be arbitrarily small.
However, in the perturbative linear UV completion, above the scale of the radial mode, this same theory is written as in \autoref{eq:higherdims} with
\begin{equation}
\begin{gathered}
\,a_0 = \frac{9}{k^4 - 10k^2 + 9}~~, \quad a_1 = 0~~,~~~~~~~~ \\
\frac{a_{n+2}}{a_n} = \frac{(2 n+7)(k^2-n^2)}{(2 n+3)(k^2-(n+5)^2)}~~,\quad a_{{n> k}} = 0~~.
\end{gathered}
\label{eq:coefs}
\end{equation}
Thus, while it appears that in the IR theory the model parameters of \autoref{eq:cosans} are $\epsilon$ and $k$, this does not at all reflect reality in the UV, where a specific tower of higher dimension operators must be generated in order to realise the IR potential of \autoref{eq:cosans}.
As a result, we see that the true UV model parameters are not simply $\epsilon$ and $k$, but actually the various $a_n$ since they correspond to the Wilson coefficients of the generated higher dimension operators.

Let us consider fine-tuning in terms of these parameters in qualitative terms.
The degeneracy of the minimum at $h/f=\pi/ (2k)$ with the other minima is a consequence of the specific values of the coefficients in \autoref{eq:coefs}.
A tiny change in one of these parameters can take the true vacuum to the minimum near $h/f = \pi(1 - 1/k)/2$.
Thus we see that a small variation in a model parameter can give rise to a large variation in the vev.
Hence the theory is fine-tuned.

Furthermore, the different RG evolution of the $a_n$'s is inevitable.
Even if the specific pattern of \autoref{eq:coefs} is generated in some UV completion of the model, the running due to the radial mode will spoil this particular pattern, giving prominence to the lower $n$ contributions and rendering the global minimum at large field values.
Or, to put it another way, to realise the potential \autoref{eq:cosans} at the matching scale requires somehow realising a different, fine-tuned set of Wilson coefficients at the deeper UV scale such that they would know to RG-evolve specifically to \autoref{eq:coefs} at the radial mode mass.

As a result, one perhaps sees most clearly through this simple perturbative model that the only radiatively stable situation in the UV is if one irrep/Gegenbauer polynomial dominates at all scales.
Only in this case can the form of the IR potential be stable against UV corrections. While this example is perturbative and employs a linearly realised symmetry in the UV, in strongly coupled UV completions the running effects will only be enhanced, exacerbating the pertinence of these aspects.

\bibliographystyle{apsrev4-1_title}
\bibliography{biblio}

\end{document}